\documentclass[a4paper,conference,twocolumn]{IEEEtran}

\usepackage[T1]{fontenc}
\usepackage[utf8]{inputenc}

\usepackage{comment}

\usepackage[cmex10]{amsmath} 
\usepackage{amssymb,amsfonts,amsthm,mathtools,physics, bbm}
\let\Set\qty
\newcommand{\field}{\mathbb{F}}

\newcommand{\eps}{\varepsilon}
\DeclareMathOperator{\supp}{supp}

\theoremstyle{remark}
\newtheorem{remark}{Remark}

\usepackage[dvipsnames]{xcolor}
\usepackage{graphicx}
\usepackage{tikz} 
\usetikzlibrary{fit}
\graphicspath{{RM_SID/}{figures/}{sfigures/}}

\makeatletter
\newcommand{\AddInputPath}[1]{%
    \providecommand*{\input@path}{}
    \g@addto@macro{\input@path}{#1}
}
\makeatother
\AddInputPath{{figures/}}
\AddInputPath{{sfigures/}}

\usepackage{hyperref}
\usepackage[capitalize]{cleveref}

\begin{document}

\title{Information-Theoretically Secret Reed-Muller Identification with Affine Designs
\thanks{We thank the DFG under Grant BO 1734/20-1, the DFG within Germany’s Excellence Strategy EXC-2111—390814868 and EXC-2092 CASA-390781972 for their support of H. Boche and M. Wiese.
Thanks also go to the BMBF within the national initiative under Grant 16KIS1003K for their support of H. Boche and under Grant 16KIS1005 for their support of C. Deppe and R. Ferrara.
We thank the research hub 6G-life under
Grant 16KISK002 for their support of H. Boche and C. Deppe.
Contact information: \{roberto.ferrara,christian.deppe,wiese, boche\}@tum.de}
}

\author{%
    \IEEEauthorblockN{%
        Mattia Spandri\IEEEauthorrefmark{1},
        Roberto Ferrara\IEEEauthorrefmark{1},
        Christian Deppe\IEEEauthorrefmark{1},
        Moritz Wiese\IEEEauthorrefmark{2}\IEEEauthorrefmark{3},
        Holger Boche\IEEEauthorrefmark{2}\IEEEauthorrefmark{3}\IEEEauthorrefmark{4}
    }
    \IEEEauthorblockA{%
        \IEEEauthorrefmark{1}Technical University of Munich, Institute for Communications Engineering, Munich, Germany\\
        \IEEEauthorrefmark{2}Technical University of Munich, Chair of Theoretical Information Technology,  Munich, Germany\\
        \IEEEauthorrefmark{3}CASA -- Cyber Security in the Age of Large-Scale Adversaries–Exzellenzcluster, Ruhr-Universit\"at Bochum, Germany\\
        \IEEEauthorrefmark{4}Munich Center for Quantum Science and Technology (MCQST), Schellingstr. 4, 80799 Munich, Germany\\
        Email: \{mattia.spandri, roberto.ferrara, christian.deppe, boche, wiese\}@tum.de
    }
}

\maketitle

    \begin{abstract}
    We consider the problem of information-theoretic secrecy in identification schemes rather than transmission schemes.
    In identification, large identities are encoded into small challenges sent with the sole goal of allowing at the receiver reliable verification of whether the challenge could have been generated by a (possibly different) identity of his choice.
    One of the reasons to consider identification is that it trades decoding for an exponentially larger rate, however this may come with such encoding complexity and latency that it can render this advantage unusable.
    Identification still bears one unique advantage over transmission in that practical implementation of information-theoretic secrecy becomes possible, even considering that the information-theoretic secrecy definition needed in identification is that of semantic secrecy.
    Here, we implement a family of encryption schemes, recently shown to achieve semantic-secrecy capacity, and apply it to a recently-studied family of identification codes, confirming that, indeed, adding secrecy to identification comes at essentially no cost.
    While this is still within the one-way communication scenario, it is a necessary step into implementing semantic secrecy with two-way communication, where the information-theoretic assumptions are more realistic.
    \end{abstract}
    
    \begin{IEEEkeywords}
    Identification, Reed-Muller codes, combinatorial designs, information-theoretic secrecy, semantic secrecy
    \end{IEEEkeywords}

\section{Introduction}
    Identification is a type of communication where the goal is verification rather than decoding.
    It can allow for either exponentially larger capacity or an exponential reduction in channel uses compared to transmission in applications where such verification suffices~\cite{AD89,AD89feedback}.
    Identification finds applications in watermarks~\cite{AC05watermarking,Moulin01,MK06}, communication complexity~\cite{Tamm01}, private interrogation~\cite{BCCK09}, monitoring in industry 4.0 and user interests in online sales and hardware stores~\cite{BD18wiretapID}, as well as
    vehicle-to-everything communication and healthcare monitoring systems~\cite{BD18wiretapID,BD19jammingID}.
    Identification has thus been proposed as one of the enabling technologies for future 6G networks~\cite{CBDSSF21}.

    In recent work~\cite{DDF20,FTBDLMV21,SFD22}, however, due to the speed of current transmission, we have hit a performance barrier in the process of implementing identification.
    Namely, current identification codes require so much processing at the sender and receiver, that the resulting encoding and decoding latency still makes transmission faster than identification, even though identification sends exponentially fewer bits.
    
    While the search for efficient identification codes is still ongoing, one reason to still use such codes is the potential of implementing information-theoretic secrecy in the wiretap channel model also at an exponentially lower cost than transmission.
    In this work, we investigate this claim at finite blocklength and show that indeed the cost of adding secrecy is small compared to the cost of using identification.
    This is done even considering that identification requires the most stringent secrecy definition of semantic secrecy~\cite{BTV12} and only few codes are explicitly proven to suffice for this criterion~\cite{HayMat, WB21, WB22}.
    Here, we use the secrecy codes of~\cite{WB22} and argue that, for identification, physical-layer secrecy might be preferable over computational secrecy.

    The paper is structured as follows.
    We define secrecy and identification in \cref{sec:preliminaries}.
    Here, we prove that Reed-Muller codes achieve noiseless identification capacity, as opposed to the opposite statement incorrectly proven in \cite{SFD22}.
    In \cref{sec:affine-designs} we describe the secrecy codes chosen from~\cite{WB22}.
    In \cref{sec:results} we present their performance, before concluding with remarks.

\section{Preliminaries}
\label{sec:preliminaries}
    We use the notation $[n]=\Set{0,...,n-1}$.
    We write
    \[
        WV(z|x) \equiv (WV)(z|x) = \sum\nolimits_{y} W(y|x) V(z|y)
    \]
    for the composition of two channels $W(y|x)$ and $V(z|y)$; $W$ could be a probability distribution in which case so is $WV$.
    We also denote by $W_x$ the probability distribution defined by $W(\cdot|x)$ for a given $x$, or $W_{Z|x}$ if the channel  has a subscript as $W_Z$.
    Analogously, if the channel has two inputs as $W(y|x,\tilde{x})$ then we can write $W_{\tilde{x}}(y|x) \equiv W(y|x,\tilde{x})$.
    We denote by
    \[
        \|P_{X}-P_{X'}\| = \sum\nolimits_{x}|P_{X}(x)-P_{X'}(x)|
    \]
    the total variation distance of two probability distributions over the same alphabet.
    With the support of $P$, denoted $\supp P$, being the set of inputs with non-zero probability, we also need the definition of {R\'{e}nyi 2-divergence} as
    \[
        D_{2}(P_{X}\|P_{X'}) = 
        \log \sum\limits_{{x \in} \supp {P_{X'}}} \frac{P_{X}(x)^{2}}{P_{X'}(x)} 
    \]
    if $\supp P_{X} \subseteq \supp P_{X'}$ and $D_{2}(P_{X}\|P_{X'}) = +\infty$ otherwise.
    We also define 
    \[
        D_{2}(W\|P_{Y}|P_{X}) = \log \sum_{x}P_{X}(x) 2^{ D_{2}(W_x\|P_Y)}.
    \]
    Finally, we denote by $P_X$ and $P_Y$ the marginals of $P_{XY}$ and by 
    \[
        I(X{:}Y) = D(P_{XY}\|P_X \cdot P_Y)
    \]
    its mutual information.

\subsection{Cryptographic vs Information-theoretic secrecy}

    Information-theoretic secrecy is well studied and understood but is little used in practice.
    Usually, practical secrecy is implemented using cryptographic secrecy based on computational assumptions (although there are now proposals for hybrid schemes~\cite{CDSM21}).
    There are two main reasons for this: the cost (which can be measured in terms of key to compare it with computational secrecy~\cite{Maurer99,Maurer12}) and the theoretical assumptions on the adversary.
    In identification, secrecy is implemented by encrypting only a marginal part of the communication.
    Thus secret identification has very low costs which nullifies the first reason against information-theoretic secrecy and also allows the use of inefficient two-way schemes~\cite{Maurer93, Maurer07} to address the second reason.

\subsection{Information-theoretic Secrecy}
\label{Chap:SeededWiretap}
\begin{figure*}
    \small
    \centering
        \begin{tikzpicture}[xscale=1.52, yscale=1,
            box/.style={draw, thick, align=center, minimum height=.5cm, minimum width=1cm}]
		\path (0,0)
    	    node[box] (source) {Source $i$}
		    ++ (1.3,0)
		    node[box] (preenc) {ID Tag\\$t_i(r)$}
		     + (0,-2) 
		    node[box] (randomness) {Uniform $r$}
		    ++ (1.2,0)
		    node[box] (encrypt) {Secrecy\\Encoder\\$h^{-1}_s(t)$}
		    ++ (0.8,0)  
		    coordinate (challenge)
		    ++ (1.1,0)
		    node[box] (enc) {Transm.\\Encoder}
		    ++ (1.2,0)
		    node[box] (channel) {Noisy\\Channel}
    		    ++ (0,-1.5)
    		    node[box, gray] (wiretap) {Wiretap\\Channel} 
    		    ++ (1.3,0)
    		    node[box, gray] (eve) {Eavesdropper}
    		    (channel)
		    ++ (1.2,0)
		    node[box] (dec) {Transm.\\Decoder}
		    ++ (1.1,0)  
		    coordinate (dechallenge)
		    ++ (0.8,0)
		    node[box] (decrypt) {Secrecy\\Decoder\\$h_s(\tilde{t})$}
		    ++ (1.2,0)
		    node[box] (postdec) {$\tilde{t} = t'${?}}
		     + (0,-2) 
		    node[box] (tagger) {ID Tag\\$t_j(\tilde{r})$}
		    ++ (1.2,0)
		    node[box] (sink) {Sink}
		    ;
        
        \draw[->] (source)      -- node[above] {$i$}                (preenc);
        \draw[->] (preenc)      -- node[above] (t-in) {$t$}         (encrypt);
        \draw[- ] (encrypt)     -- node[above] {$c$}                (challenge);
        \draw[->] (challenge)   -- node[above] {$(r,c)$}            (enc);
        \draw[->] (randomness)  -| node[below right] {$r$}          (challenge);   
        \draw[->] (enc)         -- coordinate (wire) 
                                   node[above] {$x$}                (channel);
        \draw[->] (channel)     -- node[above] {$\tilde{x}$}        (dec);
        \draw[- ] (dec)         -- node[above] {$(\tilde{r},\tilde{c})$} (dechallenge);
        \draw[->] (dechallenge) -- node[above] {$\tilde{c}$}        (decrypt);
        \draw[->] (postdec)     -- node[above] {$v$}                (sink);
        \draw[->] (randomness)  -- node[left ] {$r$}                (preenc);
        \draw[->] (sink)        |- node[below right] {$j$}          (tagger);
        \draw[->] (tagger)      -- node[right] {${t}'$}             (postdec);
        \draw[->] (decrypt)     -- node[above] (t-out) {$\tilde{t}$}(postdec);
        \draw[<-] (tagger)      -| node[below left] {$\tilde{r}$}   (dechallenge);   

        \path
            (t-in)  ++ (0,.8) node {Secret ID Challenger}
            (t-out) ++ (0,.8) node {Secret ID Verifier}
        ;

        \node [fit=(preenc)  (encrypt) (randomness), draw, dashed]{}; 
        \node [fit=(postdec) (decrypt) (tagger),     draw, dashed] {}; 

        \draw[->, gray] (wiretap) -- (eve);
        \draw[<-, gray] (wiretap) -| (wire);

        \end{tikzpicture}
        
    \caption{\label{fig:SID} Secret identification with noiseless identification codes, noiseless secrecy codes and transmission code.
    Non-secret identification is obtained by removing the secrecy encoder and decoder.
    Noiseless identification (with or without secrecy) is obtained removing the noisy channel and the transmission encoder and decoder.}
\end{figure*}
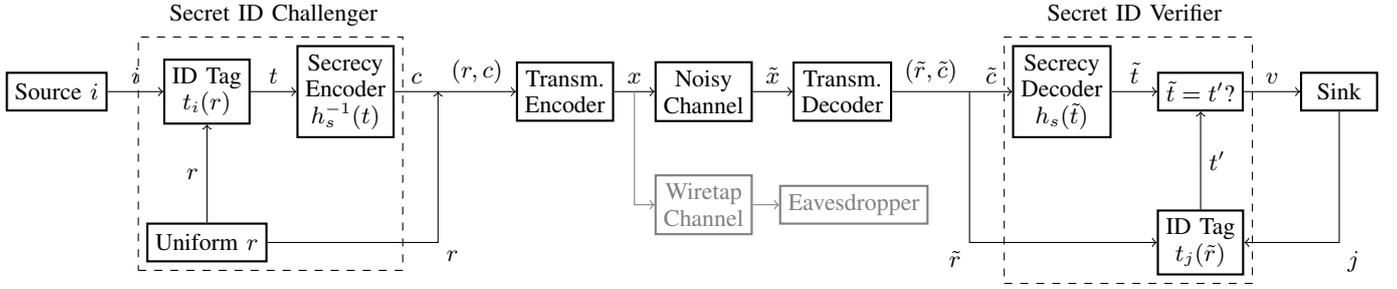

    The most general classical wiretap channel can be modelled as a joint channel $W(y,z|x)$ from the sender to the receiver and eavesdropper.
    From~\cite{Wyner75,CK78,Maurer93}, two things can be learned.
    Firstly, since the intended parties do not cooperate with the eavesdropper, the problem reduces to coding for the pair of marginal channels $W(y|x)$ and $W'(z|x)$, which is how the wiretap channel is usually modeled. 
    Secondly, it turns out that the coding for the wiretap channel separates into two codes: a first transmission code for the receiver channel to remove channel errors, which also corrects the same amount of errors at the eavesdropper, and then a second noiseless wiretap code, namely a wiretap code that assumes noiseless communication to the receiver.
    Due to this, as done in~\cite{WB22}, we assume that any needed error correction has already been performed, so that the receiver sees the input $x$ of the sender without noise, and thus we restrict without loss of generality to the case where we only have a noisy channel at the eavesdropper.
    This is further justified as identification also features a separation between the error correction of a noisy channel and identification for a noiseless one~\cite{AD89,AD89feedback}.
    This implies, as explained in more details later, that secret identification reduces to a concatenation of three codes: a noiseless identification code, a noiseless wiretap code and a noisy transmission code, where the terms noisy or noiseless are meant in the sense of ``for a noisy or noiseless channel at the receiver'' (see \cref{fig:SID}).

    A noiseless wiretap code will use a seed to map a secret message into a codeword picked at random from a set/code.
    The seed is uniform randomness that is publicly shared to the receiver, either by assuming public common randomness or by sending it to the receiver via the channel (without need for a wiretap code).
    This wiretap code is determined by a function $h:\mathcal{S} \times \mathcal{X} \to \mathcal{M}$ mapping the possible codewords $x$ and seeds $s$ to the secret messages $m = h_s(x) \equiv h(s,x)$.
    The function will be used by the receiver to decrypt $m$ from the received $x$ and $s$ while a \emph{randomized inverse} (defined next) is used by the sender together with $s$ to encrypt $m$ into $x$, which is then sent through the channel.
    The randomized inverse $h_{s}^{-1}(x|m)$ of $h_s(x)$ is defined as the channel that maps $m$ to $x$ by picking uniformly at random an element $x$ from the pre-image $h_{s}^{-1}(m) \coloneqq \{x \in \mathcal{X}|h(s,x) = m\}$.
    The sender forces the eavesdropper to make a mistake when trying to decode $z$ back to $x$ by making $h_{s}^{-1}(m)$ so large that it is beyond the capacity of the eavesdropper channel $W'(z|x)$.
    If we denote with $x'$ the decoding attempt of the eavesdropper, then the above induces an error $h(s,x') \neq m$ with high probability even though $s$ is public and known at the eavesdropper.
    Since the choice of $h$ completely defines the noiseless wiretap code, we will simply refer to $h$ as the noiseless wiretap code.

    Generally, no noiseless the wiretap code $h$ will be able to perfectly remove the information at the eavesdropper.
    Some negligible amount of information will still leak and there are various measures for this \emph{leakage}.
    One can measure the leakage in total variation, mutual information, divergences, etc, and for each of these choices one can choose between weak~\cite{Wyner75,CK78}, strong~\cite{Maurer93} or semantic secrecy~\cite{BTV12,HayMat,WB21,WB22} in order of increasing stringency.
    The secrecy capacity is the same in all measures, but a code satisfying only the weaker guarantees may allow the eavesdropper to gain too much information in some worst cases. 
    Most notably, associated with achieving strong secrecy capacity are the example of strong universal hash functions.
    Here, the requirements of identification explained in \cref{subsect:ID} below force us to use the most stringent semantic secrecy.
    Random sets, inefficient in terms of storage and computation, corresponding to unstructured $h$'s, are usually used in the achievability part of semantic secrecy capacity theorems. 
    Explicit non-random constructions with various trade-off have been presented in~\cite{HayMat,WB21}.
    We will use a family of functions from~\cite{WB22}, where some optimality in terms of encoding rate and seed size was shown.

    The semantic secrecy leakage is measured in~\cite{WB22} as either
    \begin{align}
        \label{eq:leakage}
        \max_{P_M} I(M{:}SZ) 
        & &\text{or} & &
        \max_{P_M} \norm{P_{MSZ} - P_M \cdot P_{SZ}},
    \end{align}
    where $M$ is the random variable of the message, $Z$ is the random variable of output to the eavesdropper, $S$ is the random variable for the the uniform seed independent of $M$ and where 
    \begin{gather}
        P_{SZ|M}(s,z|m) = \frac{1}{|\mathcal{S}|} h_{s}^{-1} W'(z|m)
        \\
        P_{MSZ}(m,s,z) = P_M(m)  P_{SZ|M}(s,z|m).
    \end{gather}
    ($h_{s}^{-1} W'(z|m)$ is the channel composition of $h_{s}^{-1}(x|m)$ and $W'(z|x)$).
    These are the most operational formulations of the leakage, namely they are the ones that most literally measure secrecy as ``the eavesdropper having no knowledge of my message'', which translates to the optimal distribution being ``the message is independent from the information of the eavesdropper''.
    In this paper, we will only measure the leakage as the total variation in \cref{eq:leakage}, which measures the leakage as the distance from the optimal distribution.
    Equivalent total variation measures of the leakage are also
    \begin{align}
        &\max_{m} \norm{ P_{SZ|m} - \bar{P}_{SZ} }
        &
        &\max_{m,m'} \norm{ P_{SZ|m} - \bar{P}_{SZ|m'} }
        \label{eq:indistinguishability}
    \end{align}
    with $\bar{P}_{SZ}$ being the marginal of $P_{MSZ}$ for uniform $\bar{P}_M(m) = \frac{1}{|\mathcal{M}|}$.
    These measure the maximum distinguishability of the messages from either a default average distribution or each other.
    Clearly if the output distributions are indistinguishable at the eavesdropper, then they will also be independent from the eavesdropper.

\subsection{Identification}   \label{subsect:ID}

    An identification code for $W(y|x)$ of size $I$ and error $\eps$ is a pair of challenger and verifier channels $\{C(x|i), V(v|y,i)\}$ (randomness is required for the challenger) such that they satisfy $e_{ij} = \qty|C_i W V_j(1) - \delta_{ij}|\leq \eps$, where $\delta_{ij}$ is the Kronecker delta. 
    It turns out, as mentioned already, that like for the case of secrecy, that coding for identification separates into two codes: a first transmission code for the receiver channel to remove channel errors, and then a second noiseless identification code that assumes a noiseless channel to the receiver.
    As stated before, we can thus focus only on noiseless identification codes, as done in~\cite{VW93explicit,DDF20,FTBDLMV21,SFD22}.
    In order to achieve secret identification, the noiseless wiretap code will lay between the noiseless identification code and the transmission code, as shown in \cref{fig:SID}.
    
    Let $I$ be the size a noiseless identification code and let the alphabet for the noiseless channel be $[R]\times [T] =[RT]$ for some integers $R$ and $T$.
    The capacity of noiseless identification codes is achieved by using functions to construct and verify random challenges.
    Namely, a noiseless identification code will assign to each identity $i\in[I]$ a function $t_i:[R] \to [T]$.
    The challenger $C_i$ will pick $r \in [R]$ uniformly at random (independent of the identity) and send the random challenge $(r,t) = (r,t_i(r))$, composed of the randomness $r$ and what we call the \emph{tag} $t_i(r)$.
    Finally, the verifier $V_j$ will verify the challenge by checking if the local tag is equal to the received tag; namely, it will output $v=1$ if $t_j(r) = t$ and $v=0$ otherwise.
    The errors $e_{ii}$ of such codes are always zero but the other $e_{ij}$ errors depend on the overlapping outputs between functions.
    Essentially, identification means allowing to reliably get the one bit of information at the receiver the corresponds to the question ``are the identities the same''.
    
    In secret identification~\cite{AZ95}, the goal is to not even allow this bit of information to be obtained by an eavesdropper.
    Since the only part of the communication that depends on the identity is the tag, achieving secret identification is achieved by simply by achieving secret transmission of the tag.
    Still, as the identities are compared in a pairwise manner, there is no distribution underlying the identities and thus secret identification is inherently semantic~\cite{IFD21}.
    The claim that encrypting the tag comes at asymptotically no cost follows from consideration on the size of the tag in identification-capacity achieving code.
    In the limit of identification capacity, not only the length of the challenge is exponentially smaller than the identity, but the length of the tag $t_i(r)$ itself is asymptotically zero fraction of the randomness $r$~\cite{AD89feedback} (asymptotically the tag uses zero channel capacity, while the capacity is all used to transmit $r$).
    Since the only part depending on the identity is the tag, implementing secrecy in identification comes for free asymptotically~\cite{AZ95,BD17,BD18wiretapID}.

    The large sizes that identification codes can achieve raises the question about performance of the challenger and verifier, which can introduce so much latency as to render the direct transmission of the identity $i$ to be faster than performing identification~\cite{DDF20,FTBDLMV21}.
    We choose to build on previous work, using identification codes based on Reed-Muller codes~\cite{SFD22}, the fastest identification codes known to us.
    Given a choice of prime power $q$, integers $\ell$ and $k < q$, the Reed-Muller identification code associates to each identity $i$ a multivariate polynomial $p_i$ of degree at most $k$ over $\ell$ variables in $\field_q$~\cite{KLP68,DGM70,MCJ73,SFD22}, where $i$ is the list of $\binom{\ell+k}{\ell}$ coefficients in $\field_q$ of the polynomial.
    The size of the code in bits is thus $\log I =  \binom{\ell+k}{\ell} \log q$, the challenge is composed of $\ell+1$ symbols in $\field_q$, one for the tag (thus $T=q$) and $\ell$ for the variables that are input to the polynomial (thus $R = q^\ell$).
    The resulting errors are bounded by $e_{ij} \leq \frac{k}{q}$ which is derived from the distance of the Reed-Muller code.
    \begin{remark}
    In~\cite{SFD22}, it was wrongly stated that Reed-Muller codes do not achieve identification capacity alone.
    There, $\log q \gg \log k$ was wrongly used to imply $\log k / \log q \to 0$.
    Capacity is achieved, for example, by taking the sequence of Reed-Muller identification codes with $q=2^{n^2}$, $k=2^{n^2 - n}$  and $\ell = 2^n$ and showing that they satisfy the three conditions required to achieve capacity~\cite[Eqs.~(8), (9) and~(10)]{SFD22} (coming from~\cite[Eqs.~(5), (6) and~(7)]{VW93explicit}).
    The first and third conditions a easily satisfied as $\frac{\log T}{\log R} = \frac{1}{\ell} \to 0$ and $\frac{k}{q} = 2^{-n} \to 0$.
    The second one is verified as     
    \begin{align*}
        \frac{\log\log I}{\log R} 
        &
        = \frac{\log\log q + \log \binom{k+\ell}{\ell}}{\ell \log q} 
        \to \frac{\log \binom{k+\ell}{\ell}}{\ell \log q},
        \intertext{Using the following upper and lower bounds on the binomial, $\qty(\frac{ a}{b})^b\leq\binom{a}{b}\leq \qty(\frac{\mathrm{e} a}{b})^b$ with $\mathrm{e}$ Euler's number, we bound}
        \frac{\log \binom{k+\ell}{\ell}}{\ell \log q} 
        &
        \in  \qty[\frac{\log \frac{k+\ell}{\ell}}{\log q}, \frac{\log \frac{\mathrm{e}(k+\ell)}{\ell}}{\log q} ]
        \to \frac{n^2 - 2n}{n^2} \to 1
    \end{align*}
    showing that the sequence achieves identification capacity.
    \end{remark}

\section{Secrecy via Mosaics of Affine Designs}
\label{sec:affine-designs}

\newcommand{\codeword}{x}
\newcommand{\seed}{s}
\renewcommand{\message}{m}

    In order to implement secure identification, we use the scheme depicted in \cref{fig:SID}. We assume that the channel between the sender and the verifier is noiseless, so there is no need for a transmission code. The identification code is based on Reed-Muller codes, as already mentioned in \cref{subsect:ID}. Next, we describe the secrecy functions to be applied in the scheme.
    We will restrict without loss of generality to encryption codes of rates $1/\ell' < \frac{1}{2}$ that map the tag symbol of size $q$ to multiple $\ell'$ symbols of size $q$.
    Since the tag to encrypt is such a small part of the communication, excluding encryptions that map to anything strictly between one and two symbols, especially in the case of $\ell \to +\infty$ necessary to achieve capacity, will not meaningfully impact latency or the size of the transmission.
    We will however consider the use of multiple challenges to reduce the error.
    In~\cite{SFD22}, $q$ was kept below $2^{16}$ to improve performance, but at the cost of also lower bounding the maximum error $k/q$.
    Nonetheless, this was still compared to the doubl Reed-Solomon identification codes of~\cite{DDF20}, where the error decreases exponentially, by showing that multiple challenges of Reed-Muller identification could be used to reduce the error to a similar level without impacting latency too much.
    We can thus define $\eps_{2RS}(q,k,\ell)$ as the smallest error achieved by double Reed-Solomon identification codes of size equal or below the $q$, $k$, $\ell$ Reed-Muller code.
    Here, we compute multiple Reed-Muller challenges until the identification error is below $\eps_{2RS}(q,k,\ell)$.
    Then, we encrypt each tag so that the total leakage is also comparable or below $\eps_{2RS}(q,k,\ell)$.
    This way we compare the computation time of encryption is a situation where no compromise is made on either error probability or size of the code.
    This will be explained in more details below.
    First let us define the secrecy codes and the secrecy bounds that we will use for such encryption.
    
\subsection{Secrecy codes}
    We will use the secrecy functions $h_s(x)$ induced by affine designs shown in~\cite{WB22} to provide secrecy. They are examples of secrecy functions derived from mosaics of combinatorial designs.
    Essentially, we let the codewords $x$ be elements of the vector space $\mathcal{X} = \field_q^{\ell'}$ for some length $\ell'$ that will depend on the channel and secrecy requirement.  The seed is a pair $(s, s_0) \in \mathcal{S} = \mathcal{S}' \times \field_q$, where $\mathcal{S}'$ represents the set of $(\ell'-1)$-dimensional subspaces of $\field_q^{\ell'}$ and $s_0\in\field_q^{\ell'}$.
    More precisely, $\mathcal{S}'$ is the set of vectors $s\in\field_q^{\ell'}$ with the first non-zero element normalized to one. The subspace corresponding to $s$ is the solution space of the equation $s\cdot x=0$, where $ \seed \cdot \codeword = \sum_{j} \seed_{j}\codeword_{j}$ denotes the vector inner product in $\field_q^{\ell'}$.
    The size of $\mathcal{S}'$ can be calculated with the geometric series as $|\mathcal{S}'| = \frac{q^{\ell'}-1}{q-1}$. In order to uniformly at random choose an element $s$ of $\mathcal S'$, one first chooses the index of the first non-zero entry of $s$, where the index $i$ is chosen with probability $q^{i-1}/|\mathcal{S}'|$. The entries of $s$ at positions $i-1,\ldots,1$ are then chosen uniformly at random. It is now obvious how to uniformly sample from $\mathcal{S}$.
    
    The secrecy function is constructed as follows.
    The decryption $h : \mathcal{S}' \times \field_{q} \times \field_{q}^{\ell'} \rightarrow \field_{q}$ is defined as 
    \begin{align}\label{Secrecyfunction1}
        h(\seed,\seed_0,\codeword)  = \seed \cdot \codeword + \seed_0.
    \end{align}
    For encryption, given the uniformly random $(\seed,\seed_0)$, all the components of $\codeword$ are chosen uniformly at random from $\field_{q}$ except for $\codeword_i$, where $i$ is the index of the first non-zero element of $\seed$. Then $\codeword_i$ is calculated via $\message = \seed \cdot \codeword + \seed_0$ as
    \[\codeword_{i} = \message - \seed_0 - \sum\nolimits_{j > i} \seed_{j} \codeword_j\]  
    and $\codeword$ is sent over the wiretap channel.

    In other words, given a seed $(s,s_0)$ and a secret message $m\in\field_q^{\ell'}$, the codeword to be sent is chosen from the hyperplane defined by $m-s_0 = s\cdot x$.
    Increasing the dimension $\ell'$ will increase the dimension of the hyperplanes and thus the amount of codewords in the wiretap code.

\subsection{Secrecy measure (leakage)}
    The parameters of this code as listed in~\cite[Section~2.6]{WB22} can be inserted in~\cite[Theorem~3.3~1)]{WB22} to obtain a bound on the total variation leakage of \cref{eq:leakage}:
    \begin{align*}
        &\max_{P_{M}} \norm{P_{MSZ} - P_M P_{SZ}}
        \\
        &\leq 2 \left( \frac{q^{\ell'}-q^{\ell'-1}+q^{\ell'-2}-1}{\left( q^{\ell'} -1 \right) q^{\ell'-1}}
        \left(2^{  D_{2}(W'\|P_\mathcal{X} W'|P_\mathcal{X})}-1\right) \right)^{\frac{1}{2}}
        \\&
        \leq 2 \qty(\frac{1}{q^{\ell'-1}} 2^{ D_{2}(W'\|P_\mathcal{X} W'|P_\mathcal{X})})^{\frac{1}{2}},
    \yesnumber\label{eq:leakage-upperbound}
    \end{align*}
    where $P_\mathcal{X}$ is the uniform distribution over  $\mathcal{X} = \field_q^{\ell'}$ and were we used that $q,\ell'\geq 2$ and thus $q^{\ell'}-q^{\ell'-1}+q^{\ell'-2}-1\leq q^{\ell'} -1$.

\begin{figure}
    \includegraphics[width = \columnwidth]{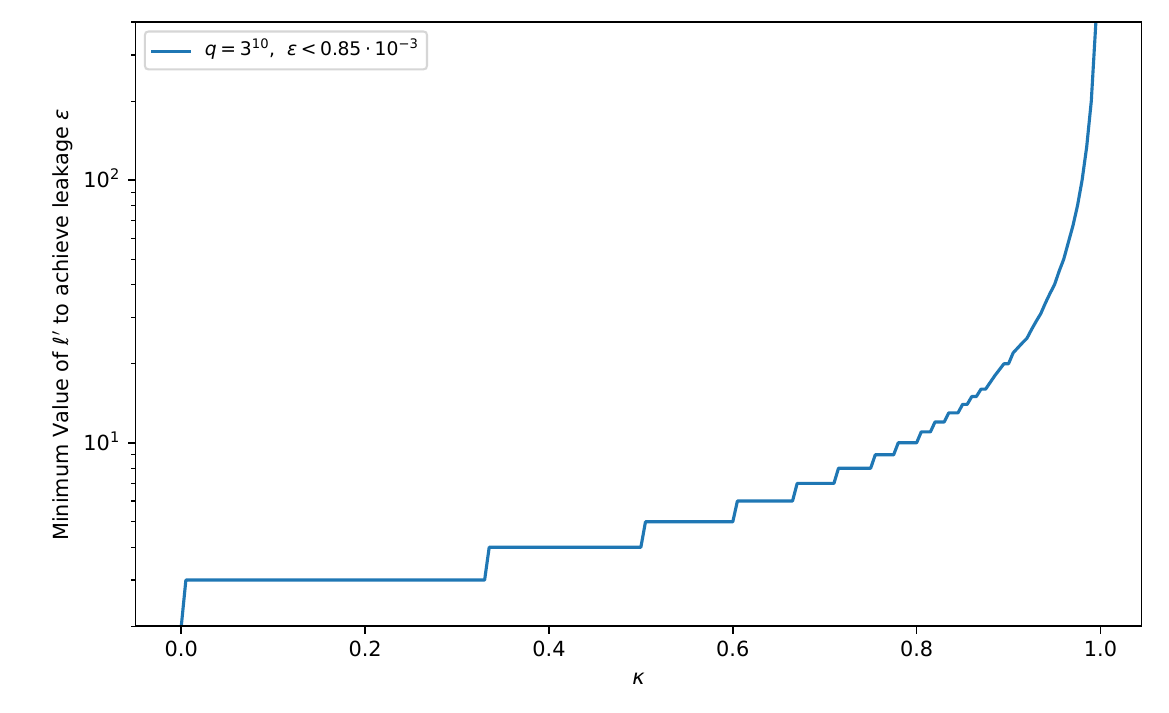}
    \caption[Minimum Parameter Value for the Mosaic of balanced incomplete block designs (BIBDs)]{The minimum number of cypher symbols $\ell'$ as a function of $\kappa$ (the channel quality of the wiretap channel) in order to achive leakage $\eps <0.85 \cdot 10^{-3}$ for symbols of size $q=3^{10}$.}
    \label{fig:Secrecy_BIBD1_t_vs_kappa}
    \end{figure}

    This is the point where the secrecy depends on the wiretap channel and thus on the assumption on the eavesdropper.
    We cannot expand the above upper bounds further without knowing the channel $W'$.
    However, recall that 
    \begin{align}
        0 \leq D_{2}(W'\|P_{\mathcal{X}}W'|P_{\mathcal{X}}) \leq \ell' \log q.
    \end{align}
    We can thus parametrize the $D_2$ measure of the channel as a number of bits of information bounded by $\ell' \log q$ or, as we will do, as a fraction $\kappa$ of the bits of information sent.
    Then we have
    \begin{align}
        D_{2}(W'\|P_{\mathcal{X}}W'|P_{\mathcal{X}}) \equiv q^{\kappa \ell'}.
    \end{align}
    Now, suppose that we want the leakage to be bounded by $\eps$.  In terms of the above equation, this means
    \begin{align}
        \max_{P_{M}} \norm{P_{MSZ} - P_M P_{SZ}} 
        &
        \leq 2 \qty( \frac{q}{ q^{(1-\kappa)\ell'}} )^{\frac{1}{2}}
        \leq \eps
        .
    \end{align}
    Taking the logarithm, this amounts to
    \begin{align*}
        1 + \frac{1}{2} (\log q - (1-\kappa)\ell' \log q) \leq \log \eps.
    \end{align*}
    {From this we can now obtain a lower bound for $\ell'$ that guarantees that the leakage will be upper bounded by $\eps$:}
    \begin{align}
        \ell' \geq \frac{2 + \log q + 2 \log \eps}{(1-\kappa) \log q}
        \label{eq:cypher-minimum-length}
    \end{align}
    An example of such lower bound is displayed in \cref{fig:Secrecy_BIBD1_t_vs_kappa}.

\subsection{Secret Identification}
    As shown by \cref{eq:cypher-minimum-length}, in order to decide the length of our encryption, we need to decide what kind of secrecy level we want to achieve, namely we have to decide how much information we are comfortable leaking to the adversary as measured by $\eps$.
    The choice of secrecy parameter $\eps$ depends on the application and we thus have to consider the identification codes we want to use and come up with an appropriate level of secrecy.
    The first choice we make is to require that the leakage be of the same order or lower than the maximum error $\max_{ij} e_{ij}$ of the identification code.
    The rationale is that the eavesdropper will anyway not be able to distinguish whether a wrong identification was induced by a failed eavesdropping or a failed identification.
    Notice that this is markedly different from computational secrecy, where a small probability of failure to encrypt may mean being able to recover the whole message.
    Here, even if the eavesdropper does everything right, it will only obtain an amount of information bounded by $\eps$.
    
\begin{figure}
        \centering
        \includegraphics[width = \columnwidth]{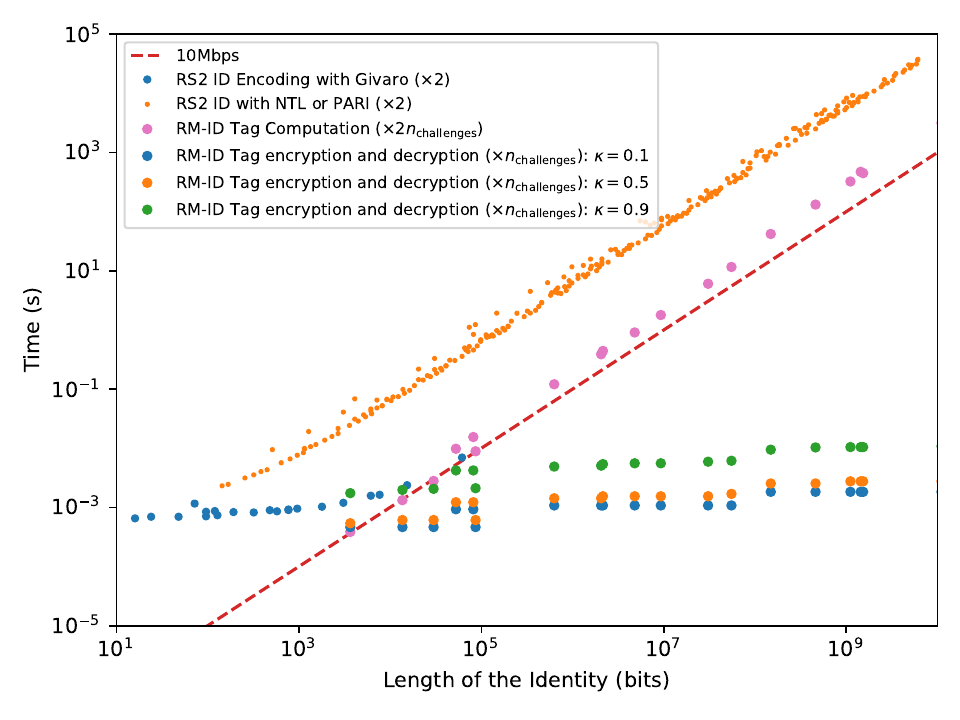}
        \caption[Latency of Identification Codes and Secrecy Codes]{Computation time of various parts of secret identification. The orange (and dark blue) points are the times of challenge generation for the code from~\cite{DDF20}, which provides the secrecy requirement $\eps$.
        The red line would be the time for a single challenge generation for the fastest Reed-Muller identification codes from~\cite{SFD22} which we encrypt, but multiple challenges are computed in order to bring the identification error down $\eps$. 
        This increases slightly the slope as seen comparing the pink points and the red line.
        The remaining points are the tag encryption times for different values of the fraction $\kappa$ of information bits at the eavesdropper.}
        \label{fig:Encoding_and_Secrecy_time_rtimesm_multi_challenges}
    \end{figure}

    We want to encrypt the Reed-Muller identification codes. However, as remarked in~\cite{SFD22}, with a single challenge and bounded $q$, the maximum error does not decrease as the size increases. Instead, multiple challenges must be used for that.
    At the same time, we expect to reduce the error to zero as we increase the code size.
    Thus for reference, we need a capacity-achieving code to determine the decreasing-error rate so that the number of challenges can be chosen in the Reed-Muller identification code.
    As anticipated, for this, we can take the double Reed-Solomon identification code from~\cite{VW93explicit,DDF20} which was both the motivation and the comparison for the Reed-Muller identification code.
    We also pick from~\cite{SFD22} the subset of fastest parameters of the Reed-Muller codes for a given code size, which are achieved for $k \gg \ell$, giving the set of parameters $q=3^{10}$, $\ell \in [2,6]$ and $k \in \{10,20,30,40,50\}\cdot \ell$.
    Recall then that we defined $\eps_{2RS}(q,k,\ell)$ as the smallest error achieved by double Reed-Solomon identification codes of size equal or below the $q$, $k$, $\ell$ Reed-Muller code.
    Since we need to encrypt multiple $n$ challenges, we want the joint encryption to have leakage bounded by $\eps_{2RS}(q,k,\ell)$ which we can guarantee by asking each challenge to leak at most $\eps_{2RS}(q,k,\ell)^\frac{1}{n}$.
    The size of the encryption codes $\ell'$ is then obtained by inserting this maximum leakage in \cref{eq:cypher-minimum-length}.
    Again, \cref{fig:Secrecy_BIBD1_t_vs_kappa} shows an example of how the length $\ell'$ scales with increasing eavesdropping fraction $\kappa$, displaying the expected blow up to infinity for $\kappa\to 1$.
    \Cref{fig:Encoding_and_Secrecy_time_rtimesm_multi_challenges} shows the resulting encoding time of the multiple challenges and the resulting encryption and decryption time for this amount of challenges with the number of symbols needed to achieve $\eps_{2RS}$.

\section{Computation Time and Discussion}
\label{sec:results}
    The encryption and decryption were implemented in Python and SageMath~\cite{Sage}, since the Reed-Muller identification code was also implemented with the same language and libraries.
    We measured the time of tag computation, encryption and decryption in a server equipped with two Intel\textsuperscript{\textregistered} Xeon\textsuperscript{\textregistered} E5-2687W v4 running Arch Linux.
    The resulting extra time needed for encryption for a few values of the wiretap parameter $\kappa$ is shown in \cref{fig:Encoding_and_Secrecy_time_rtimesm_multi_challenges}.
    As seen by the almost constant graphs, the small size of the tag permits the use of information-theoretic encryption at no cost of latency compared to the time needed to produce the identification challenges, as expected.
    However, the size of the identities larger than $10^4$ to $10^5$ are needed for a clear cut advantage over transmission codes.
    Such sizes might be largely unnecessary if the identity is used to identify a certain number of devices/users, but reasonable if used to identify data.
    Future work will be directed into implementing two-way secrecy which allows for more realistic assumptions.

\section*{Acknowledgments}
    The authors acknowledge the financial support by the Federal Ministry of Education and Research
    of Germany (BMBF) in the programme of “Souverän. Digital. Vernetzt.”. Joint project 6G-life, project identification number: 16KISK002.
    H. Boche and M. Wiese were further supported in part by the BMBF within the national initiative on Post Shannon Communication (NewCom) under Grant 16KIS1003K.
    C.D. and R.F. were funded by the Bavarian State Ministry for Economic Affairs, Regional Development and Energy in the project 6G and Quantum Technology (6GQT).
    C. Deppe was also supported by the DFG within the project DE1915/2-1.

\bibliographystyle{alpha}

\begin{thebibliography}{10}
\csname url@samestyle\endcsname
\providecommand{\newblock}{\relax}
\providecommand{\bibinfo}[2]{#2}
\providecommand{\BIBentrySTDinterwordspacing}{\spaceskip=0pt\relax}
\providecommand{\BIBentryALTinterwordstretchfactor}{4}
\providecommand{\BIBentryALTinterwordspacing}{\spaceskip=\fontdimen2\font plus
\BIBentryALTinterwordstretchfactor\fontdimen3\font minus
  \fontdimen4\font\relax}
\providecommand{\BIBforeignlanguage}[2]{{%
\expandafter\ifx\csname l@#1\endcsname\relax
\typeout{** WARNING: IEEEtran.bst: No hyphenation pattern has been}%
\typeout{** loaded for the language `#1'. Using the pattern for}%
\typeout{** the default language instead.}%
\else
\language=\csname l@#1\endcsname
\fi
#2}}
\providecommand{\BIBdecl}{\relax}
\BIBdecl

\bibitem{AD89}
R.~Ahlswede and G.~Dueck, ``Identification via channels,'' \emph{IEEE
  Transactions on Information Theory}, vol.~35, no.~1, pp. 15--29, 1989.

\bibitem{AD89feedback}
------, ``Identification in the presence of feedback-a discovery of new
  capacity formulas,'' \emph{IEEE Transactions on Information Theory}, vol.~35,
  no.~1, pp. 30--36, 1989.

\bibitem{AC05watermarking}
R.~Ahlswede and N.~Cai, \emph{Watermarking Identification Codes with Related
  Topics on Common Randomness}.\hskip 1em plus 0.5em minus 0.4em\relax Berlin,
  Heidelberg: Springer Berlin Heidelberg, 2006, pp. 107--153.

\bibitem{Moulin01}
P.~Moulin, ``The role of information theory in watermarking and its application
  to image watermarking,'' \emph{Signal Processing}, vol.~81, no.~6, pp.
  1121--1139, 2001.

\bibitem{MK06}
P.~Moulin and R.~Koetter, ``{A framework for the design of good watermark
  identification codes},'' in \emph{Security, Steganography, and Watermarking
  of Multimedia Contents VIII}, vol. 6072.\hskip 1em plus 0.5em minus
  0.4em\relax SPIE, 2006, pp. 565 -- 574.

\bibitem{Tamm01}
U.~Tamm, \emph{Communication complexity and orthogonal polynomials}.\hskip 1em
  plus 0.5em minus 0.4em\relax American Mathematical Society, 2001, vol.~56,
  pp. 277--285.

\bibitem{BCCK09}
J.~Bringer, H.~Chabanne, G.~Cohen, and B.~Kindarji, ``Private interrogation of
  devices via identification codes,'' in \emph{Progress in Cryptology -
  INDOCRYPT 2009}, B.~Roy and N.~Sendrier, Eds.\hskip 1em plus 0.5em minus
  0.4em\relax Berlin, Heidelberg: Springer Berlin Heidelberg, 2009, pp.
  272--289.

\bibitem{BD18wiretapID}
H.~{Boche} and C.~{Deppe}, ``Secure identification for wiretap channels;
  robustness, super-additivity and continuity,'' \emph{IEEE Transactions on
  Information Forensics and Security}, vol.~13, no.~7, pp. 1641--1655, 2018.

\bibitem{BD19jammingID}
------, ``Secure identification under passive eavesdroppers and active jamming
  attacks,'' \emph{IEEE Transactions on Information Forensics and Security},
  vol.~14, no.~2, pp. 472--485, 2019.

\bibitem{CBDSSF21}
J.~Cabrera, H.~Boche, C.~Deppe, R.~F. Schaefer, C.~Scheunert, and F.~H.~P.
  Fitzek, ``{6G and the Post-Shannon-Theory},'' in \emph{{Shaping Future 6G
  Networks: Needs, Impacts and Technologies}}, T.~M. Emmanuel~Bertin,
  Noel~Crespi, Ed.\hskip 1em plus 0.5em minus 0.4em\relax Wiley-Blackwell,
  2021.

\bibitem{DDF20}
S.~Derebeyoğlu, C.~Deppe, and R.~Ferrara, ``Performance analysis of
  identification codes,'' \emph{Entropy}, vol.~22, no.~10, p. 1067, 2020.

\bibitem{FTBDLMV21}
R.~Ferrara, L.~Torres-Figueroa, H.~Boche, C.~Deppe, W.~Labidi, U.~M\"onich, and
  V.-C. Andrei, ``Practical implementation of identification codes,'' 2021.

\bibitem{SFD22}
M.~Spandri, R.~Ferrara, and C.~Deppe, ``Reed-muller identification,'' in
  \emph{International Zurich Seminar on Information and Communication (IZS
  2022). Proceedings}, A.~Lapidoth and S.~M. Moser, Eds.\hskip 1em plus 0.5em
  minus 0.4em\relax Zurich: ETH Zurich, 2022, pp. 74 -- 78.

\bibitem{BTV12}
M.~Bellare, S.~Tessaro, and A.~Vardy, ``Semantic security for the wiretap
  channel,'' in \emph{Advances in Cryptology -- CRYPTO 2012}, R.~Safavi-Naini
  and R.~Canetti, Eds.\hskip 1em plus 0.5em minus 0.4em\relax Berlin,
  Heidelberg: Springer Berlin Heidelberg, 2012, pp. 294--311.

\bibitem{HayMat}
M.~Hayashi and R.~Matsumoto, ``Secure multiplex coding with dependent and
  non-uniform multiple messages,'' \emph{{IEEE} Transactions on Information
  Theory}, vol.~62, no.~5, pp. 2355--2409, 2016.

\bibitem{WB21}
M.~Wiese and H.~Boche, ``Semantic security via seeded modular coding schemes
  and ramanujan graphs,'' \emph{IEEE Transactions on Information Theory},
  vol.~67, no.~1, pp. 52--80, 2021.

\bibitem{WB22}
------, ``Mosaics of combinatorial designs for information-theoretic
  security,'' \emph{Designs, Codes and Cryptography}, vol.~90, no.~3, pp.
  593--632, Mar. 2022.

\bibitem{CDSM21}
A.~Cohen, R.~G.~L. D’Oliveira, S.~Salamatian, and M.~Médard, ``Network
  coding-based post-quantum cryptography,'' \emph{IEEE Journal on Selected
  Areas in Information Theory}, vol.~2, no.~1, pp. 49--64, 2021.

\bibitem{Maurer99}
U.~Maurer, ``Information-theoretic cryptography,'' in \emph{Advances in
  Cryptology --- CRYPTO' 99}, M.~Wiener, Ed.\hskip 1em plus 0.5em minus
  0.4em\relax Berlin, Heidelberg: Springer Berlin Heidelberg, 1999, pp. 47--65.

\bibitem{Maurer12}
------, ``Constructive cryptography -- a new paradigm for security definitions
  and proofs,'' in \emph{Theory of Security and Applications},
  S.~M{\"o}dersheim and C.~Palamidessi, Eds.\hskip 1em plus 0.5em minus
  0.4em\relax Berlin, Heidelberg: Springer Berlin Heidelberg, 2012, pp. 33--56.

\bibitem{Maurer93}
------, ``Secret key agreement by public discussion from common information,''
  \emph{IEEE Transactions on Information Theory}, vol.~39, no.~3, pp. 733--742,
  1993.

\bibitem{Maurer07}
\BIBentryALTinterwordspacing
U.~Maurer, R.~Renner, and S.~Wolf, \emph{Unbreakable Keys from Random
  Noise}.\hskip 1em plus 0.5em minus 0.4em\relax London: Springer London, 2007,
  pp. 21--44. [Online]. Available:
  \url{https://doi.org/10.1007/978-1-84628-984-2_2}
\BIBentrySTDinterwordspacing

\bibitem{Wyner75}
A.~D. Wyner, ``The wire-tap channel,'' \emph{The Bell System Technical
  Journal}, vol.~54, no.~8, pp. 1355--1387, 1975.

\bibitem{CK78}
I.~Csisz\'ar and J.~K\"orner, ``Broadcast channels with confidential
  messages,'' \emph{IEEE Transactions on Information Theory}, vol.~24, no.~3,
  pp. 339--348, 1978.

\bibitem{VW93explicit}
S.~{Verdu} and V.~K. {Wei}, ``Explicit construction of optimal constant-weight
  codes for identification via channels,'' \emph{IEEE Transactions on
  Information Theory}, vol.~39, no.~1, pp. 30--36, 1993.

\bibitem{AZ95}
R.~Ahlswede and Z.~Zhang, ``New directions in the theory of identification via
  channels,'' \emph{IEEE Transactions on Information Theory}, vol.~41, no.~4,
  pp. 1040--1050, 1995.

\bibitem{IFD21}
A.~Ibrahim, R.~Ferrara, and C.~Deppe, ``Identification under effective
  secrecy,'' in \emph{2021 IEEE Information Theory Workshop (ITW)}, 2021, pp.
  1--6.

\bibitem{BD17}
H.~Boche and C.~Deppe, ``Robust and secure identification,'' in \emph{2017 IEEE
  International Symposium on Information Theory (ISIT)}, 2017, pp. 1539--1543.

\bibitem{KLP68}
T.~Kasami, S.~Lin, and W.~Peterson, ``New generalizations of the reed-muller
  codes--i: Primitive codes,'' \emph{IEEE Transactions on Information Theory},
  vol.~14, no.~2, pp. 189--199, 1968.

\bibitem{DGM70}
P.~Delsarte, J.-M. Goethals, and F.~J. Mac~Williams, ``On generalized
  reed-muller codes and their relatives,'' \emph{Information and control},
  vol.~16, no.~5, pp. 403--442, 1970.

\bibitem{MCJ73}
J.~L. Massey, D.~J. Costello, and J.~Justesen, ``Polynomial weights and code
  constructions,'' \emph{IEEE Transactions on Information Theory}, vol.~19,
  no.~1, pp. 101--110, 1973.

\bibitem{Sage}
T.~S. Developers. Sagemath, the sage mathematics software system (version 8.1),
  2020.

\end{thebibliography}


\end{document}